# *Bioinformatics* Supplementary Material

February 14, 2009

## S1. Application to plant cell culture data

## Replicated microarray time course

Modeling of causal networks has advanced significantly over the last decade using simple eukaryotic systems. For example, approaches using synchronized yeast cells sampled at short intervals and profiled for gene expression changes have led to the development of gene regulatory models that provide useful information about the cell cycle (e.g. Bickel 2005; Spellman, Sherlock, Zhang, Iyer, Anders, and others 1998).

To determine the extent to which influences between genes can be inferred from a microarray time course in a system relevant to crop improvement, we designed an experiment using a plant cell suspension system (BMS) that has a history of success as a simple cell system for biochemical and gene expression studies. Cells in the stationary stage were perturbed by the addition of a single hormone treatment and sampled at 10-minute time intervals over 150 minutes for measurement of gene expression changes by transcriptional profiling. We chose abscisic acid (ABA) because it has relevance to the plant's response to abiotic stresses such as drought (low water availability) and cold (low temperature) and because ABA is a relatively well-characterized inducer of gene expression (Verslues and Zhu 2005). Based on previous experiments, we identified a list of twenty-five ABA response genes that showed a consistent induction in the BMS cell culture system by five hours following ABA treatment. Determining which genes regulate as many of these twenty-five genes of interest as possible provides evidence for stress-relevant parts of the gene network of the system.

The rest of this section describes the data analysis results, gene annotations, and details of the experimental and preprocessing methods used in preparation for the statistical methods specified above.

## Influences between genes

### Application of kinetic model inference

Figure S1 displays the expression intensities of all 25 genes that were pre-selected as regulated by genes to be identified by the models given the data. The posterior probabilities were computed using approximations (8), (12), and (14) for each of 25 genes of interest as regulated by some unknown gene with expression more directly changed by the ABA treatment. Table S1 reports marginal probabilities of each difference-equation order and what may be interpreted as each interest gene's probability that its regulating gene could be identified on the basis of the expression data. The predictions, residuals, and intensities of the probable regulating genes of the four of the 25 genes for which the first approximation achieved 50% probability are plotted in Figures S2-10. In Figures S2-6, the "probability" is $P\left(\alpha_{ij} = 1 \middle| \mathbf{y}\right)$ of equation (8), whereas in Figures S7-10, it is $P\left(\tilde{\alpha}_{ij} = 1 \middle| \mathbf{y}\right)$ of equation (12). Figure S11 quantifies the extent to which the posterior probabilities (14) are sensitive to the joint prior distribution of the models and parameters represented by equation (13).

| Regulated gene of interest | Marginal posterior probability of each order over all genes | | Probability that the highest-probability gene is regulated by the interest gene (maximized separately for each column) | | | |
|---|---|---|---|---|---|---|
| | Probability that order 1 | Probability that order 2 | 1st-order conditional | 2nd-order conditional | 1st- & 2nd-order models | |
| ID229128 | 100.0% | 0.0% | 94.6%* | 30.6% | 92.5% | a |
| ID263329 | 100.0% | 0.0% | 76.5%* | 59.8% | 73.3% | b |
| ID225167 | 97.6% | 2.4% | 62.6%* | 99.5% | 56.9% | c |
| ID295057 | 84.8% | 15.2% | 58.7%* | 68.8% | 45.6% | d |
| ID273437 | 99.7% | 0.3% | 39.0% | 44.9% | 35.8% | e |
| ID272371 | 98.8% | 1.2% | 39.4% | 47.5% | 34.0% | f |
| ID295063 | 48.8% | 51.2% | 42.1% | 56.5% | 29.4% | g |
| ID269778 | 65.0% | 35.0% | 19.8% | 80.6% | 28.9% | h |
| ID221767 | 75.5% | 24.5% | 6.3% | 99.8% | 25.1% | i |
| ID227718 | 98.1% | 1.9% | 25.1% | 17.6% | 19.5% | j |
| ID232009 | 99.9% | 0.1% | 18.5% | 52.8% | 14.8% | k |
| ID246472 | 88.4% | 11.6% | 21.3% | 66.2% | 13.9% | l |
| ID297343 | 98.7% | 1.3% | 17.3% | 17.7% | 13.9% | m |
| ID281792 | 100.0% | 0.0% | 16.0% | 18.8% | 12.6% | n |
| ID277341 | 93.7% | 6.3% | 12.4% | 28.5% | 8.9% | o |
| ID290708 | 99.8% | 0.2% | 10.9% | 8.3% | 8.8% | p |
| ID287048 | 91.6% | 8.4% | 7.0% | 79.5% | 6.9% | q |
| ID248343 | 72.8% | 27.2% | 10.0% | 25.4% | 6.5% | r |
| ID221443 | 100.0% | 0.0% | 11.1% | 15.6% | 6.5% | s |
| ID246471 | 99.6% | 0.4% | 5.4% | 75.1% | 4.8% | t |
| ID274390 | 99.8% | 0.2% | 4.8% | 46.8% | 3.9% | u |
| ID240829 | 95.7% | 4.3% | 4.9% | 9.7% | 3.4% | v |
| ID239234 | 99.2% | 0.8% | 3.1% | 16.1% | 2.4% | w |
| ID296385 | 99.9% | 0.1% | 1.9% | 20.0% | 1.3% | x |
| ID273436 | 98.4% | 1.6% | 0.4% | 1.2% | 0.3% | y |

**Table S1. Posterior probabilities summarizing the data analysis results.** The first two numeric columns report (15) and (16) for each gene given in the leftmost column and presumed regulated by some dominant regulating gene to be identified by the analysis. The last three numeric columns give the posterior probability of the best-fitting model (corresponding to the most probable regulating gene) for each of the 25 genes of interest assumed to be regulated by a gene responding more directly to ABA treatment. Those three maximum posterior probability columns, from left to right, were computed using the models of equations (7), (11), and (13); the last percentage column best represents our state of uncertainty in choosing between the assumptions behind the other two columns (cf. Figure S11). Details on the genes with a regulating gene of at least 50% posterior probability in the first-order column (*) are given in Figures S2-10. The data and models provide no reason to infer any other components of the gene network unless one has reason apart from this experiment to strongly favor assumptions leading to the second-order models; Figures S7-10 are given for that case with the same genes included.

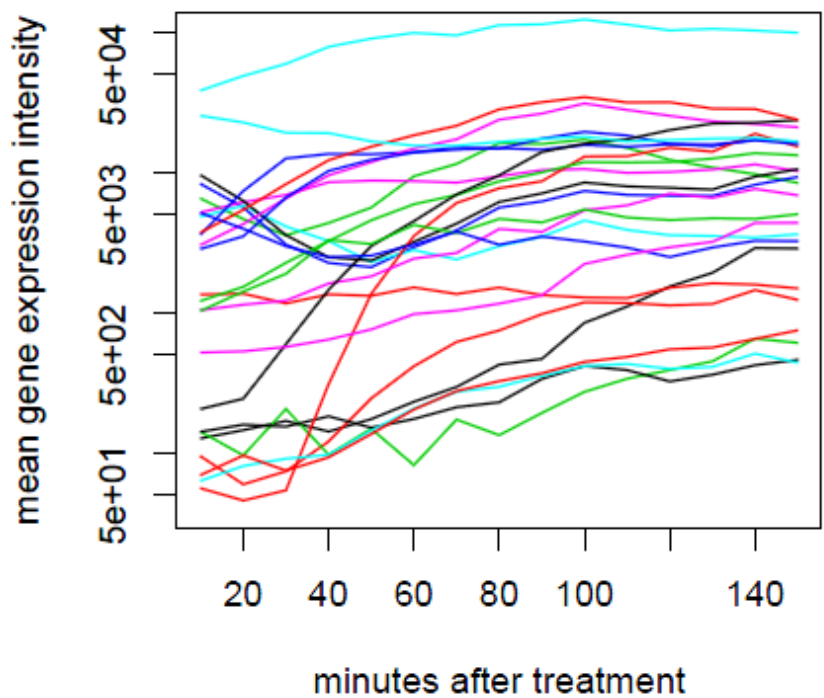

**Figure S1.** Means of observed expression intensities of the 25 genes of interest, each assumed to be regulated by a gene that responded more directly to the ABA treatment.

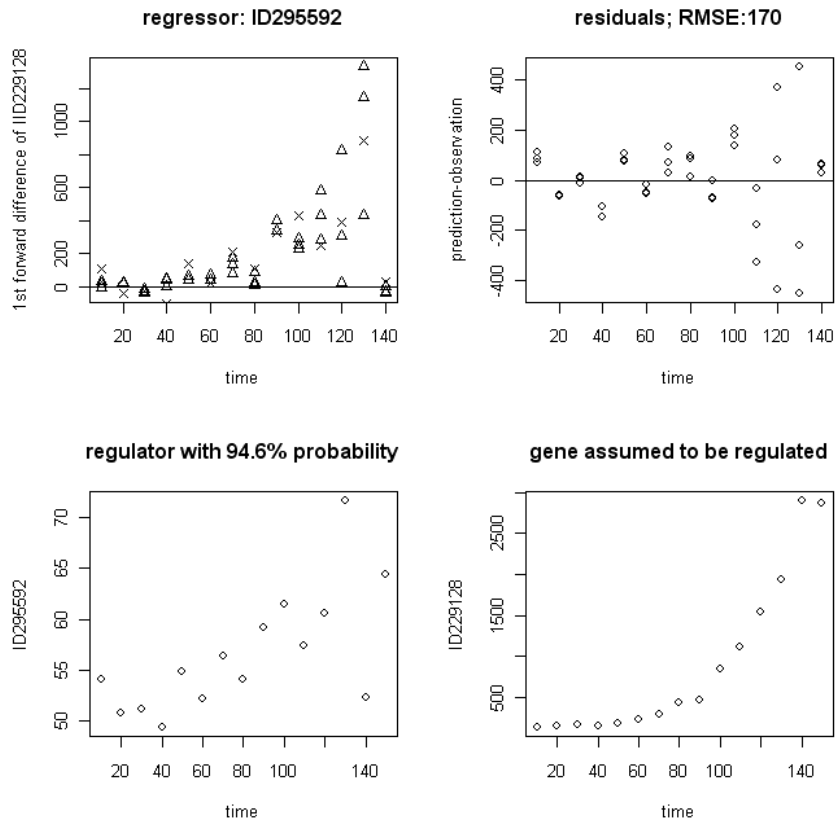

**Figure S2.** Of the 25 genes of interest assumed regulated by at least one regulating gene that responded more directly to the ABA treatment, ID229128 ("a" in Table S1) had a gene, ID295592, with the highest posterior probability of being the regulator. The model fit is quantified in the top two plots; the left plot displays $\Delta y_{ik}(t)$ (symbol: $\Delta$) and $\beta_{ij}\overline{y}_j(t) + \beta_i$ (symbol: $\times$), whereas the right plot displays $\varepsilon_{ik}(t)$. The bottom two plots give $\overline{y}_j(t)$ for each of the two genes, assisting with the physical interpretation: until 120 minutes after the addition of ABA, the regulating gene steadily accumulates mRNA, triggering the regulated gene to produce mRNA at an increasing rate; at $t = 130$ minutes, the concentration of the regulating gene's transcript is markedly higher, resulting in the jump in that of the regulated from 130 minutes to 140 minutes; finally, the move of the regulating gene at 140 minutes back to its initial level halts the transcript production of the regulated gene, seen to be the same at 150 minutes as at 140 minutes.

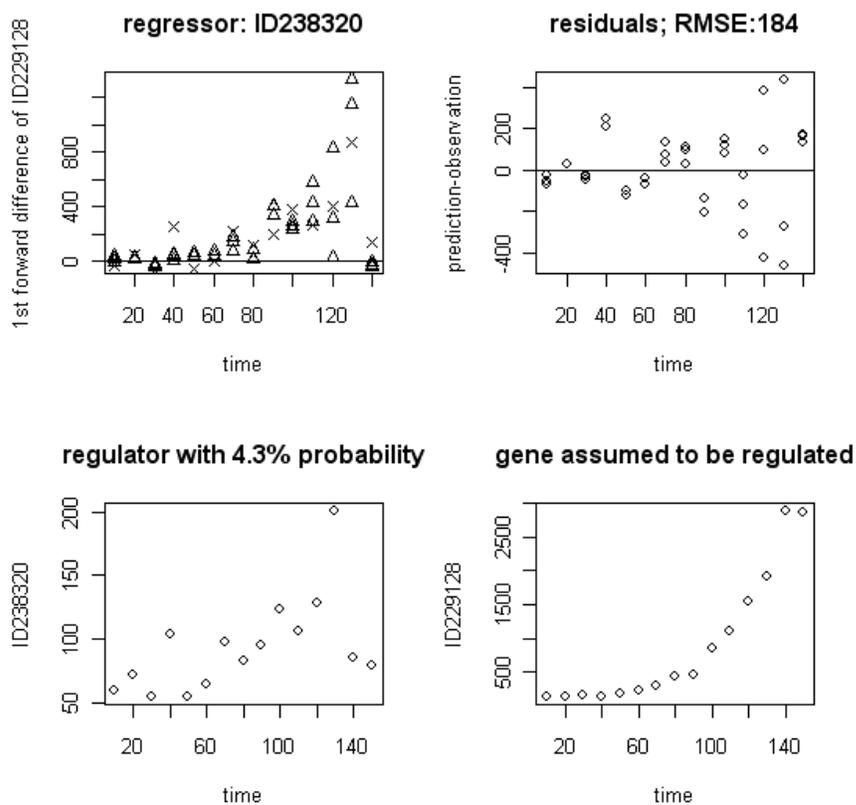

**Figure S3.** The regulated gene was chosen because its potentially regulating gene of *highest* probability (Figure S2) exceeded 50%, but with the potentially regulating gene of *second highest* posterior probability shown here. In spite of the much lower model probability (4.3%), the fit here does not look much worse, except at 140 minutes after treatment: the transcript of the regulating gene does not fall all the way back to its initial value.

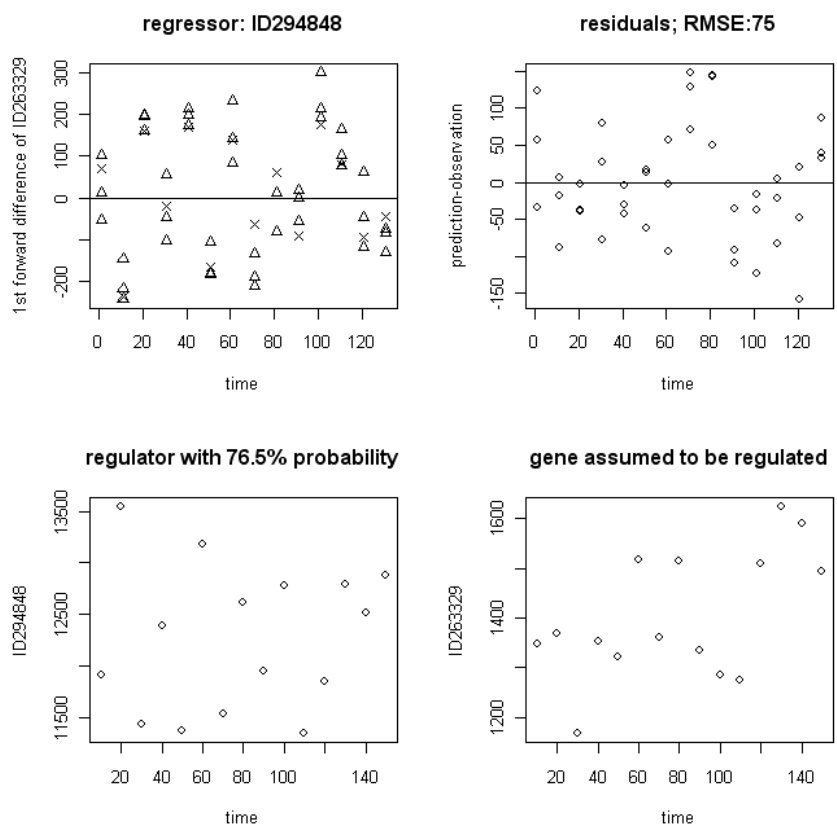

**Figure S4.** The assumed regulated gene here ("b" in Table S1), among the genes of interest (Table S1), has a regulating gene with the *second* highest model posterior probability. The sudden jumps in the rate of change of the regulated gene are well explained by the earlier jumps in the regulating gene, which, unlike that of Figure S2, appears to be a repressor. However, the time scale of these fluctuations appears to be smaller than that in a cell (Kristin Baetz and Mads Kaern, personal communication), perhaps indicating a need to incorporate additional information into the model.

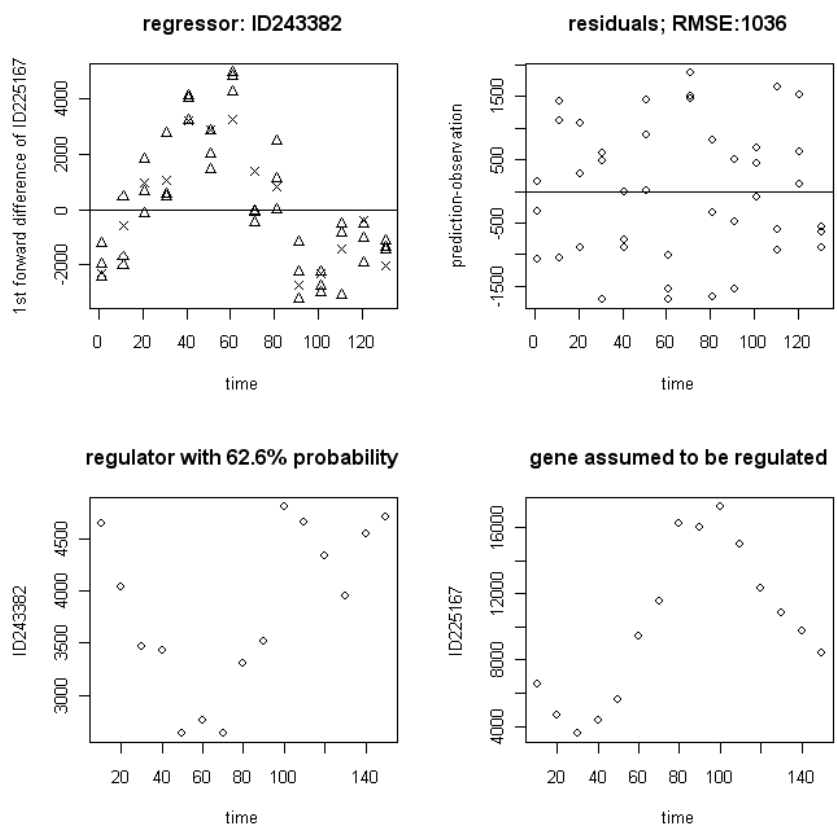

**Figure S5.** The assumed regulated gene here ("c" in Table S1), among the genes of interest (Table S1), has a regulating gene with the *third* highest model posterior probability. The physical interpretation here would again be that the regulating gene represses the regulated gene.

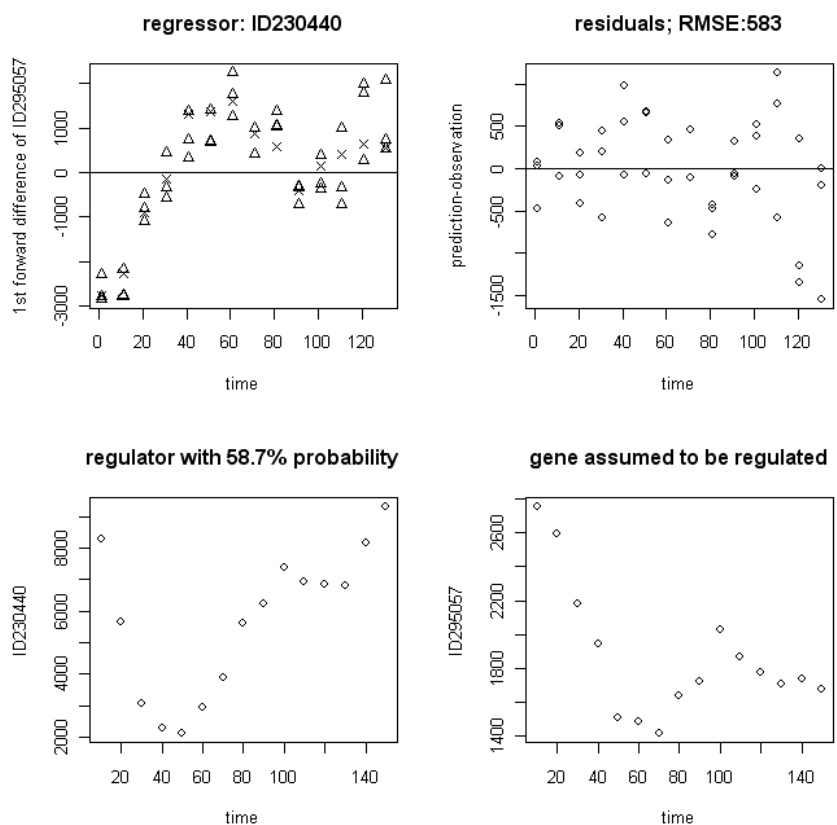

**Figure S6.** The assumed regulated gene here ("d" in Table S1), among the genes of interest (Table S1), has a regulating gene with the *fourth* highest model posterior probability. That posterior probability is just over half, so the activation relationship is about as likely as not according to the mathematical framework. The physical interpretation here would again be that the regulating gene represses the regulated gene.

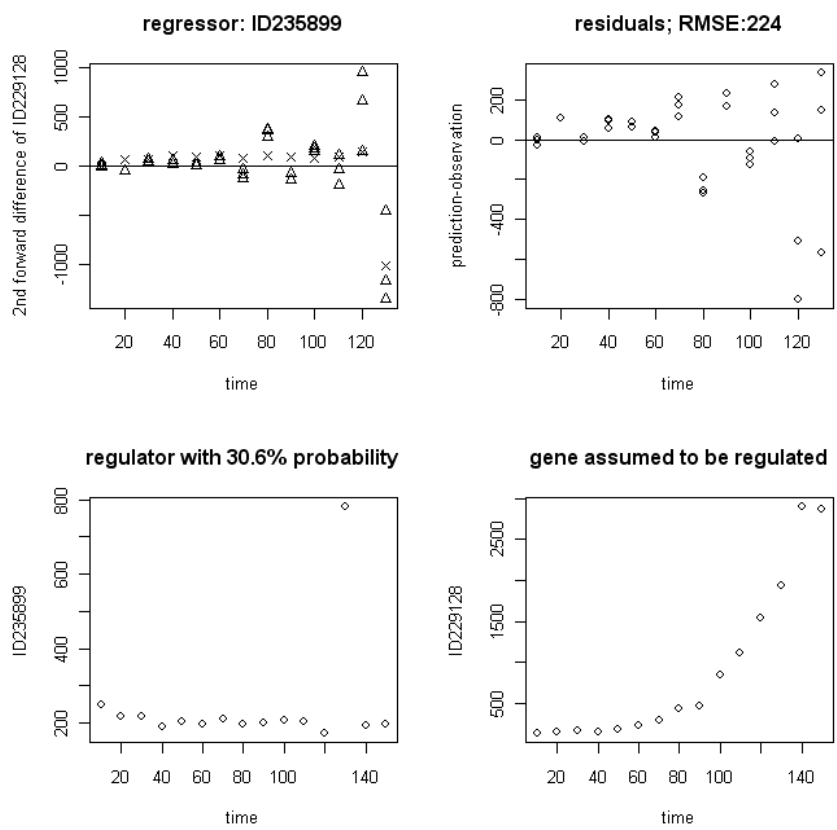

**Figure S7.** The best second-difference fit for the regulated gene of Figure S2 ("a" in Table S1).

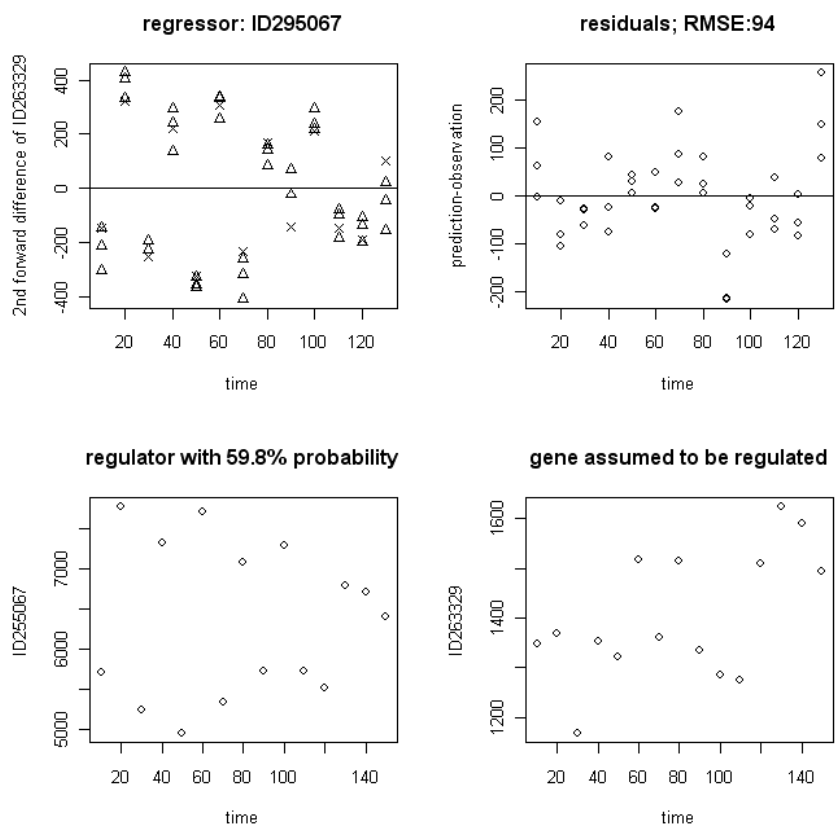

**Figure S8.** The best second-difference fit for the regulated gene of Figure S4 ("b" in Table S1).

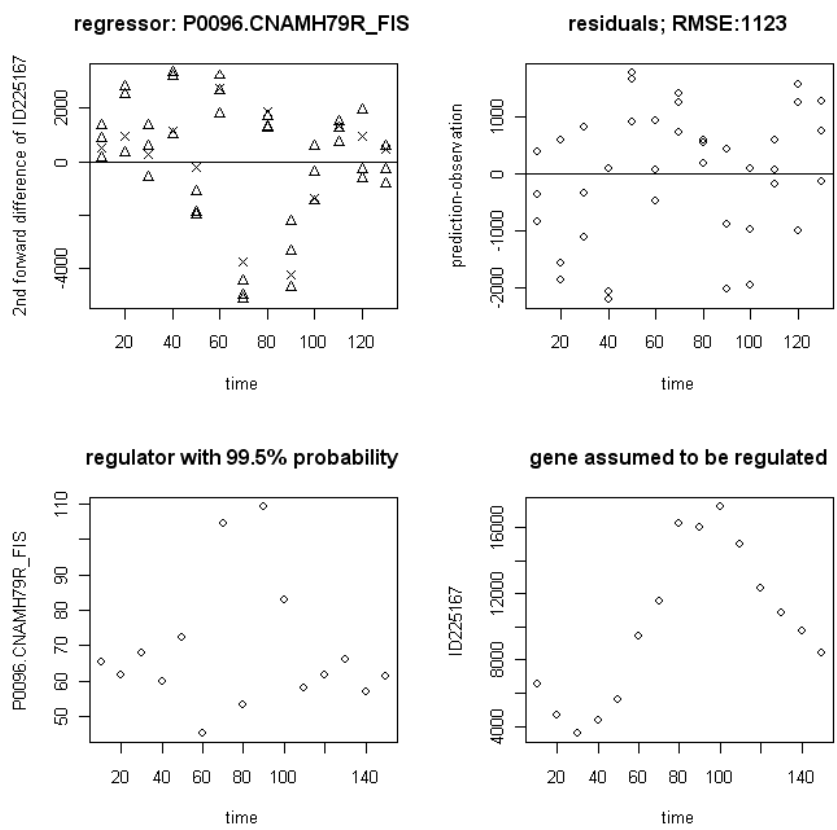

**Figure S9.** The best second-difference fit for the regulated gene of Figure S5 ("c" in Table S1).

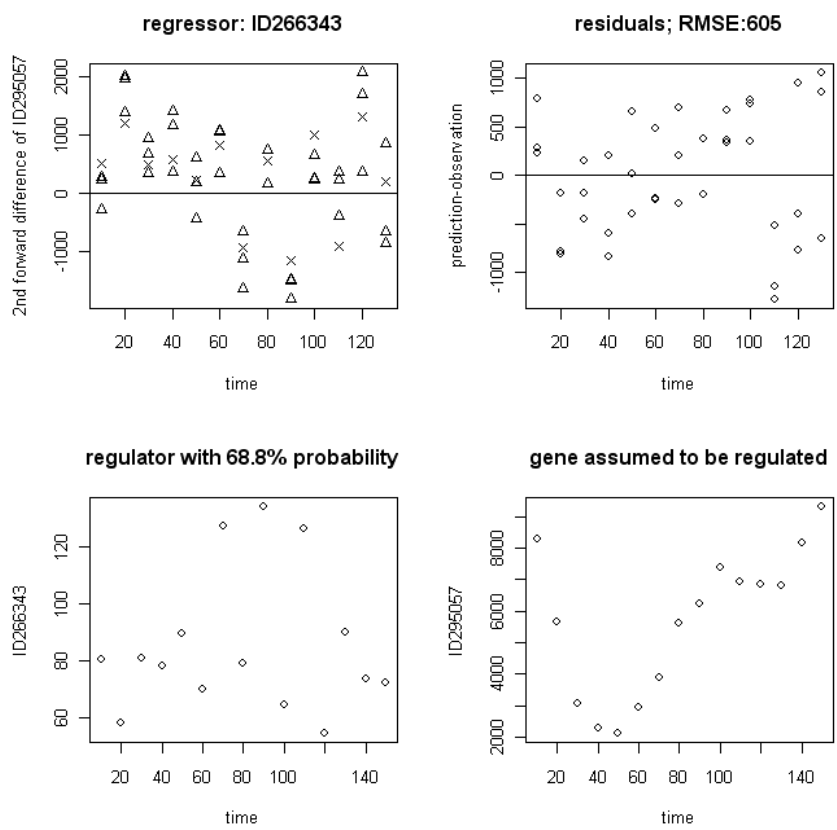

**Figure S10.** The best second-difference fit for the regulated gene of Figure S6 ("d" in Table S1).

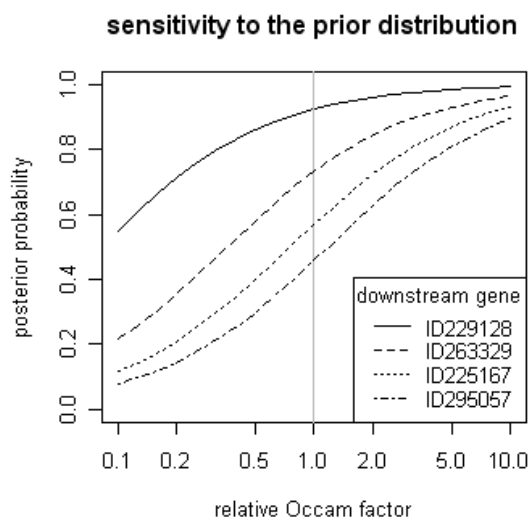

**Figure S11.** The posterior probability versus the relative Occam factor for each of the regulated genes of Figures S2-10. As explained in Section S3, the relative Occam factor measures departures of the prior distribution from the default, which has a relative Occam factor of 1. Here, a gene is predicted to regulate gene $i$ if it maximizes, over all genes, the posterior probability that it is the regulator of that regulated gene of interest. That maximum probability,

$$\max_{j \in \{1,\ldots,m\}} P\left(\alpha_{ij} = 1 \cup \tilde{\alpha}_{ij} = 1 \,\middle|\, \mathbf{y}\right),$$

is displayed here as a function of the relative Occam factor $\Omega_i$, which in turn depends on the choice of the conditional prior distributions of the regression coefficients. A plot versus the relative prior mass $\omega_i$ instead of the relative Occam factor would look exactly the same according to Section S3. The vertical line corresponds to the values found in Table S1 under "**1st- & 2nd-order models**."

## Gene annotations for the probable interactions

Agilent profiling is based on a 60-mer array derived from EST sequence information. We used the 60-mer oligonucleotide for each of the tags representing the individual genes to BLAST against public databases and attempt to establish biological relevance to some of the genes identified in the modeling process. Table S2 summarizes the BLAST output for each of the eight genes involved in probable interactions according to the first-order difference model.

ID229128 represents a gene with homology to cold-regulated proteins and is a member of a family of stress and ABA responsive genes with homology to phosphatidylethanolamine binding proteins (Dal Bosco, Busconi, Govoni, Baldi, Stanca, Crosatti, Bassi, and Cattivelli 2003). The biochemical function of this protein, as it relates to ABA and abiotic stress, has not been characterized but this gene is induced 13.4 fold following 5 hours of ABA treatment in the BMS system (data not shown). ID295592 presumably regulates ID229128 and represents a gene tag with no significant homology to genes of known function and weak homology to proteins that function as transposable elements. Further characterization of ID295592 would be required to understand its functional role in stress and ABA signaling.

ID263329 corresponds to ABA hydroxylase, an enzyme involved in the conversion of ABA into 8'-hydroxyl catabolite products. The gene that probably regulates ID263329 corresponds to a TAB2 protein homolog (ID294848). TAB2 is an RNA binding protein that functions to enhance chloroplast gene translation. ID225167 and its presumed regulating gene (ID243382) do not have significant homology to proteins of known function. Lastly, ID295057 corresponds to a NAC1 transcription factor, a large gene family with members that are associated with stress tolerance (Kikuchi and Hirano 2000). The gene predicted to regulate ID295057 corresponds to a phospatidylinositol-4-phosphate 5-kinase-like gene which has been previously shown to be involved in ABA signaling (Mikami, Katagiri, Iuchi, Yamaguchi-Shinozaki, and Shinozaki 1998).

| Gene | GenBank Reference | BLASTX Annotation | BLASTX Score |
|------|-------------------|-------------------|--------------|
| ID295592 | CN844307.1 | ABA95172.1\|  transposon protein, putative | 1 |
| ID229128 | BM382127 | CAC12881 cold-regulated protein | 8E-33 |
| ID294848 | DV023912.1 | BAD19229.1 putative Tab2 protein | 6E-80 |
| ID263329 | BZ736934.1 | ABB71585.1\| ABA 8'-hydroxylase 1 | 1E-100 |
| ID243382 | AZM5_88405 | ABF93726.1 Unknown protein (Oryza sativa) | 4E-08 |
| ID225167 | AY109653.1 | NP_910324.1\| unknown protein | 4E-24 |
| ID230440 | AZM5_1672 TIGR | BAD38030.1\| 1-phosphatidylinositol-4-phosphate 5-kinase-like protein | 2E-61 |
| ID295057 | CV071472 | ABD52007 stress-induced transcription factor NAC1 | 2E-18 |

**Table S2.** Annotations of the genes plotted in Figures S2 and S4-10. Individual Agilent 60-mer oligonucleotide tags were BLASTed against public databases to arrive at a GenBank reference gene. Shaded rows correspond to putative regulators, and bold borders separate each putative regulating-regulated pair.

## Case study methods

### Design of the microarray time-course experiment

*Sample collection.* Sixty *Zea mays* Black Mexican Sweet (BMS) cell suspension of 40 ml in medium 237 (De Rocher, Vargo-Gogola, Diehn, and Green 1998) were pooled and centrifuged at 10,000xg for 15 minutes. These cells were washed once in 500 ml of 237 lacking 2,4-D and resuspended in the same medium. The cells divided into 45 ml cultures in 250 ml Erlenmeyer flasks and grown for 48 h at 28ºC shaking at 160 rpm on an orbital shaking platform under dark conditions to allow for acclimation to hormone-free medium. To normalize the samples against any uncontrolled variability in culture conditions, the cells were pooled again and well mixed before redistribution into 51 aliquots of 40 ml each in identical 250 ml Erlenmeyer flasks. Six of these cultures were harvested on glass fiber filters on an analytical vacuum filter at time 0 minutes as control samples, three of which were used in the expression profiling. Each of the remaining 45 cultures was treated with a final concentration of 34μM ABA (Sigma Biosciences, St. Louis) and sampled by filtering between 10 and 150 minutes later, such that three cultures were sampled at each 10-minute interval. Thus, three biological replicates of each time point were collected, resulting in 3 reference samples and 45 treatment samples from 10 to 150 minutes after the addition of ABA. All sampling was performed on the same day with every three treated cultures of the same number of ABA minutes sampled at approximately the same astronomical time.

*Gene expression profiling.* For the 48 samples (consisting of 3 at time zero and three at every 10 minutes thereafter), a total of 90 slides were hybridized: there were 45 comparisons to time 0 and a technical replicate (dye swap) for each comparison. Total RNA was isolated from frozen ground tissue (Purescript, Gentra System) followed by polyA RNA isolation (mRNA Purification kit, Amersham Biosciences). Samples were then amplified and labeled (Cy3 and Cy5) using Agilent's Low RNA Linear Amp kit and purified with Qiagen's RNeasy column. Each labeled sample was hybridized to a 44k Agilent microarray with one of three reference samples (the other three were held in reserve) measured at time 0, when ABA was added. Dye swaps were set up for all comparisons to account for any dye bias effects to give a total of 90 slides hybridized. The microarray slides were hybridized overnight, washed, and scanned according to Agilent's Two-Color Microarray-Based Gene Expression Analysis protocol. Images were analyzed with Agilent's Feature Extraction Software (v 9.1) and visually inspected for image artifacts. Further quality control analysis was done using data analysis tools in Rosetta's Resolver database software. Technical replicates (dye swaps) were combined, generating a weighted average of the individual values. The probes on the slides correspond to 42,044 putative genes (expressed sequence tags), referred to herein as "genes." Since the anticipated number of genes in the *Zea mays* genome is of the same order of magnitude, we use $m = 42,044$. Intensity data are available from http://www.oisb.ca/downloads.htm.

## Data preprocessing

Whereas the differential equations considered model the dynamics of transcript concentrations in a cell, data available from microarrays only give rough measures of the total mRNA copy number in all cells of each sample taken from the cell culture or tissue of interest. Accordingly, we make the assumption that the microarray intensities to some extent reflect cell concentrations up to the second derivative, an assumption that would be justified if, for example, the mRNA copy numbers in some synchronous cells of each sample are an order of magnitude greater than those of the sample's cells that have different expression dynamics. A more thorough approach would model the relationship between single-cell dynamics and the resulting microarray intensities to the extent that informative prior distributions would be placed on the number of cells per sample and on aspects of interactions between cells. Although such an approach would better quantify uncertainty and potentially lead to very different conclusions, it poses significant mathematical and computational difficulties. Research to bridge the gap between biochemically informed differential equations and the scale of expression currently measured by microarrays is currently underway (Wilkinson 2006).

With that caveat in mind, the observed transcript concentrations, $y_{jk}(t)$, were computed up to an irrelevant proportionality constant as follows. After averaging intensities over multiple probes corresponding to the same sequences, the two-color Agilent microarray referenced by replicate $k$ and time $t$ was hybridized to two cell culture samples, yielding two intensity values, $y_{jk}^{+}(t)$ and $y_{jk}^{-}(t)$, for gene $j$. The "+" superscript corresponds to a sample extracted at time $t$ after the addition of ABA, whereas the "−" superscript corresponds to a reference sample extracted at time 0; the argument "$t$" in $y_{jk}^{-}(t)$ only indicates the time of the sample hybridized to the same slide. The simplest approach assuming intensity-concentration proportionality would set $y_{jk}(t)$ equal to $y_{jk}^{+}(t)$, thereby discarding the information in the reference samples. However, to mitigate the effect of microarray-specific variability, we instead applied the correction

$$y_{jk}(t) = y_{jk}^{+}(t) - \left[ y_{jk}^{-}(t) - \sum_{k'=1}^{3} y_{jk'}^{-}(t)\Big/3 \right].$$

In the event of zero variance between reference samples, the terms enclosed by square brackets cancel each other, achieving the same result as the simplest approach. This correction also has the desirable property that it does not affect the mean over all replicates of the same time: $\sum_{k''=1}^{3} y_{jk''}(t)\Big/3 = \sum_{k''=1}^{3} y_{jk''}^{+}(t)\Big/3$.

## Hypothesis generation

Gene expression profiling is increasingly applied as a tool for understanding complex biological systems. Despite the capability to accurately measure changes in expression of a large number of genes through profiling, our ability to extract applicable knowledge lags behind, largely due to difficulties in reliably inferring gene interactions as well as a lack of functional understanding for individual genes. These limitations are addressed in part by the development of statistical methodology for inferring networks of genes on the basis of high-throughput expression data.

We selected ABA-response genes for modeling that had little previous characterization in an attempt to build knowledge by association irrespective of known biological function. For four of those twenty-five regulated genes, each of their main regulating genes could be identified on the basis of the expression data with at least 50% probability according to the first-order difference model.

Two of the interactions predicted by this study identified unknown and uncharacterized genes (ID295592 and ID243382) each as the regulator of a largely unknown-function target gene (ID229128 or ID225167, respectively). Since the biochemical function of these genes is not characterized, an understanding of their interaction will require further study. With sufficient additional experiments, a network map could be developed that would eventually anchor these genes to known biological processes.

For the interaction ID263329/ID294848, the regulating gene is modeled as a repressor of the regulated ABA hydroxylase. This is interesting in light of the fact that ID294848 corresponds to a chloroplast translation factor and RNA binding protein. RNA binding proteins have been implicated in ABA function (Sabine, Zsolt, Christian de, Thomas, and Stefan 2006) and the chloroplast is the site of ABA biosynthesis (Qin and Zeevaart 1999). Taken together, this putative interaction presents a testable hypothesis that a feedback mechanism exists between ABA biosynthesis and degradation and that ABA may play a role in signaling from the chloroplast, as has recently been suggested (Shen, Wang, Wu, Du, Cao, Shang, Wang, Peng, Yu, Zhu, Fan, Xu, and Zhang 2006).

Signal transduction intermediates in the ABA response pathway have been identified by genetic analysis and by gene expression studies (Rock 2000), and recently two ABA receptors have been identified and characterized (Razem, El-Kereamy, Abrams, and Hill 2006; Shen, Wang, Wu, Du, Cao, Shang, Wang, Peng, Yu, Zhu, Fan, Xu, and Zhang 2006). However, a linear path from ABA to response gene has not been fully elucidated and the issue of cross-talk with other hormone pathways remains contentious. Gene network analysis tools presented in this manuscript and in development provide a distinct approach from mutant and gene expression analysis that may serve to clarify the complexity in how genes interact following a hormone stimulus. Supporting this idea, our analysis identified a transcription factor from the NAC family (ID295057) that appears to be regulated by a regulating gene involved in inositol triphosphate signaling (PIPK), which has been demonstrated to play a role in ABA signaling (Lin, Ye, Ma, Xu, and Xue 2004; Mikami, Katagiri, Iuchi, Yamaguchi-Shinozaki, and Shinozaki 1998). The identification of ID230440 as a regulating repressor of ID295057 is particularly significant since PIPK is thought to be involved in re-setting the IP3 signal (Mikami, Katagiri, Iuchi, Yamaguchi-Shinozaki, and Shinozaki 1998), whereas NAC genes have been characterized as positive regulators of the ABA signal (Kikuchi and Hirano 2000).

## Computation time

In the 32-bit Windows implementation of R version 2.7.1 with a 2.3 GHz processor and 3 GB of RAM, it takes 6 minutes to compute the 42,044 model-averaged posterior probabilities corresponding to each gene of interest or 150 minutes for all 25 genes of interest. Thus, seeking regulators of all 42,044 genes would require about 6 months on that platform. The use of C or another low-level language instead of R could make such genome-wide scans feasible with similar hardware.

# S2. Application to bacteria and yeast data sets

## Bacteria and yeast data sets

To quantify the performance of our models, we chose four sets of public data: two bacteria data set (Kao *et al*., 2004; Bansal *et al*., 2006) and two yeast data sets (Spellman *et al.,* 1998; de Lichtenberg *et al.*, 2005). All of the data have been normalized by the authors.

| | Data set | |
|---|---|---|
| | **Kao *et al*., 2004** | **Bansal *et al*., 2006** |
| **Strain** | BW25113 | MG1655 |
| **Time interval** | 12 min | 1 h |
| **Number of time points** | 6 | 6 |
| **Number of genes** | 4345 | 4290 |
| **Data condition** | Missing data | Complete data |

**Table S3.** Attributes of the two *E. coli* data sets**.**

| | Data set | |
|---|---|---|
| | **Spellman *et al*., 1998** | **de Lichtenberg *et. al.*, 2005** |
| **Strain** | DBY8724 | CDC28-13 |
| **Time interval** | 7 min | 10 min |
| **Number of time points** | 18 | 17 |
| **Number of cell cycles** | 2 | 2 |
| **Number of ORFs** | 6178 | 6214 |
| **Data condition** | Missing data | Complete data |

**Table S4.** Attributes of the two *S. cerevisiae* data sets**.** The open reading frames (ORFs) in the two yeast data sets include ORFs currently considered uncharacterized or dubious as well as those considered verified.

## The genes of interest

From a search of the literature, we culled a set of genes reported to be regulated to serve as the targets of interest. In total, 55 *E. coli* genes and 43 *S. cerevisiae* genes were selected. The genes are categorized into 12 groups on the basis of reasons for selection; the genes and their groups are listed in the supplementary Excel spreadsheet "results.xls".

The *E. coli* genes appear in five groups labelled B1 to B5. Group B1 contains 11 LexA-regulated genes in the SOS pathway (Quillardet, *et al*., 2003). (The SOS pathway is an important regulatory network found in many bacteria that mediates the response to DNA damage.) Group B2 contains genes which are considered highly upregulated by RpoS, that is, genes for which the mean expression ratio is greater than 4 with P-values under 0.001 (Patten, *et al*., 2004). Group B3 contains genes from the same paper which are considered downregulated using the equivalent criterion to the genes in Group B2. Groups B4 and B5 contain upregulated and downregulated genes, respectively, following an acid shift (Kannan, *et al*., 2008).

The *S. cerevisiae* genes of interest are categorized into seven groups (labelled Y1 to Y7). The genes in the first three groups were identified as strongly cell cycle-regulated genes in Spellman *et al*. (1998). This study used a statistic called "Fourier score" to estimate how closely the periodicity in the expression of each gene follow matched the period of the cell cycle. High Fourier scores were assumed to indicate cell-cycle-coupled regulation. P-values were calculated using a reference distribution created by randomly permuting real profiles and calculating Fourier scores for the resulting artificial profiles. Groups Y1 to Y3 contain genes assigned as regulated with decreasing confidence. Genes in Group Y1 have Fourier scores greater than 10 with p-values below 0.05 and false discovery rates (FDRs) below 10%. The genes in Group Y2 have high Fourier scores for which the p-values are still below 0.05 but for which the FDRs are between 10% and 40%. Group Y3 contains genes having Fourier scores such that the p-values are between 0.05 and 0.2 and the FDRs are below 20%.

The genes of Group Y4 are considered strongly regulated in Spellman *et al*. (1998) but not in de Lichtenberg *et al*. (2005). Group Y5 contains four cell cycle-regulated genes which are induced by alpha factor (Spellman *et al*., 1998). In Group Y6, we included two weakly expressed cell cycle-regulated genes (de Lichtenberg *et al.,* 2005) to test the ability of our model to predict not just strongly regulated genes but also weakly regulated genes. Finally, the three genes in Group Y7 are hypothesized to be related to each other on the basis of a nonparametric method of reconstructing gene networks (Bickel, 2004).

## Prediction results

For each target gene considered regulated, three posterior probabilities were separately maximized over all genes: the posterior probabilities of the first-order model (7), the second-order model (11), and the average model (13), thereby finding, for each model, the gene with the highest probability of regulating each gene of interest. The annotations of regulating genes were checked in EchoBASE, GeneDB, and the *Saccharomyces* Genome Database for *E. coli* genes and *S. cerevisiae* genes.

We found how many of the genes that maximize the posterior probability are also putative transcription factors (reported in Figures S12-15). These comparisons indicate that the second-order model and average model are better able to predict transcription factors than the first-order model since most of the putative transcription factors have posterior probabilities greater than 0.5 in the second-order model and the average model.

Figures S12-15 show the first-order, the second-order, and the average models for each data sets. These figures indicate that the number of putative transcription factors in the *E. coli* data sets is more than in the *S. cerevisiae* data sets, and also that the regulating genes of *E. coli* have higher posterior probabilities than those in *S. cerevisiae*. These results may reflect the fact that *S. cerevisiae* gene networks are more complex than those of *E. coli*.

Information on the genes of Figures 1-3 and S12-15 is available in the supplementary Excel spreadsheet "results.xls." Some entries in the spreadsheet are marked by "NA" to indicate that one of the regulated genes (YMR307W) in the *S. cerevisiae* list is not included in the de Lichtenberg *et al.* (2005) data set. The spreadsheet also gives the estimates of coefficients $\beta_{ij}$ and $\tilde{\beta}_{ij}$ for the genes of highest posterior probabilities.

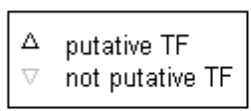

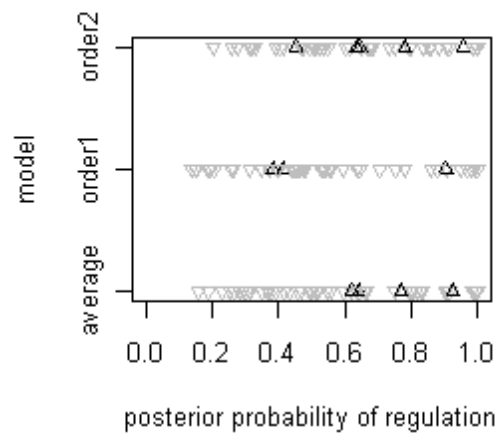

**Figure S12.** Results for the *E. coli* data set from Kao *et al.* (2004). The posterior probabilities of regulation for the average (equation (14)), first-order (equation (8)), and second-order (equation (12)) models are shown as "average," "order1," and "order2," respectively. Within a given model, each triangle represents a probability-maximizing gene that corresponds to a different gene of interest. Black triangles represent probability-maximizing genes listed in the EchoBASE as putatively encoding transcription factors (TFs), and gray triangles denote other probability-maximizing genes.

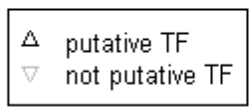

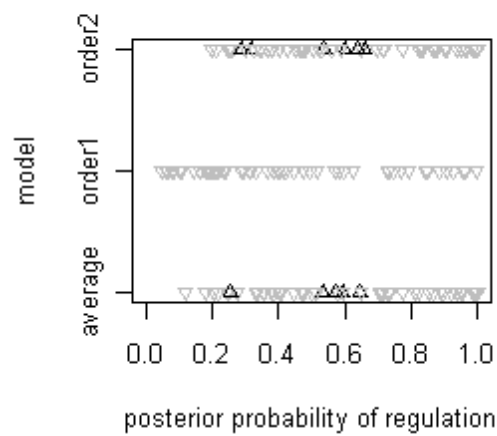

**Figure S13**. Results for the *E. coli* data set from Bansal *et al.,* 2006. Figure labels and symbols are as in Figure S12.

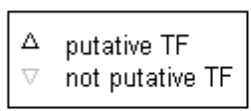

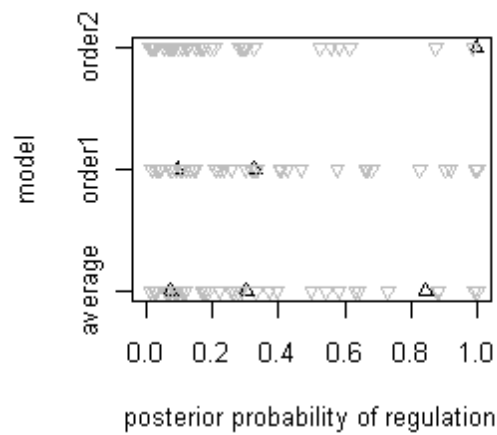

**Figure S14.** Results for the *S. cerevisiae* data set from Spellman *et al.,* 1998. Figure labels and symbols are as in Figure S12.

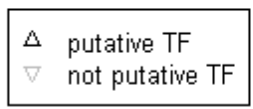

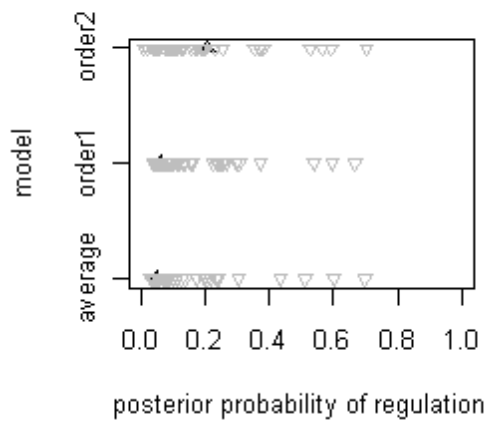

**Figure S15.** Results for the *S. cerevisiae* data set from de Lichtenberg *et al*., 2005. Figure labels and symbols are as in Figure S12.

Table 1 of the main text reports estimates of the Area Under the receiver operating characteristic Curve (AUC) computed by

$$\frac{1}{n_{TF}n_{NTF}}\left(\left|\left\{(i,j):i\in\{1,2,..,n_{TF}\},j\in\{1,2,..,n_{NTF}\},x_i>y_j\right\}\right|\right),$$

where $x_i$ is the posterior probability associated with the $i^{th}$ of $n_{TF}$ putative transcription factors and $y_j$ is the $j^{th}$ of $n_{NTF}$ genes that are not transcription factors. The AUC is the frequentist probability that the posterior probability of a gene randomly chosen from the population of putative transcription factors is greater than the posterior probability of a gene randomly chosen from the population of genes that are not putative transcription factors. Figure S16 displays the estimated AUCs for each of the four data sets.

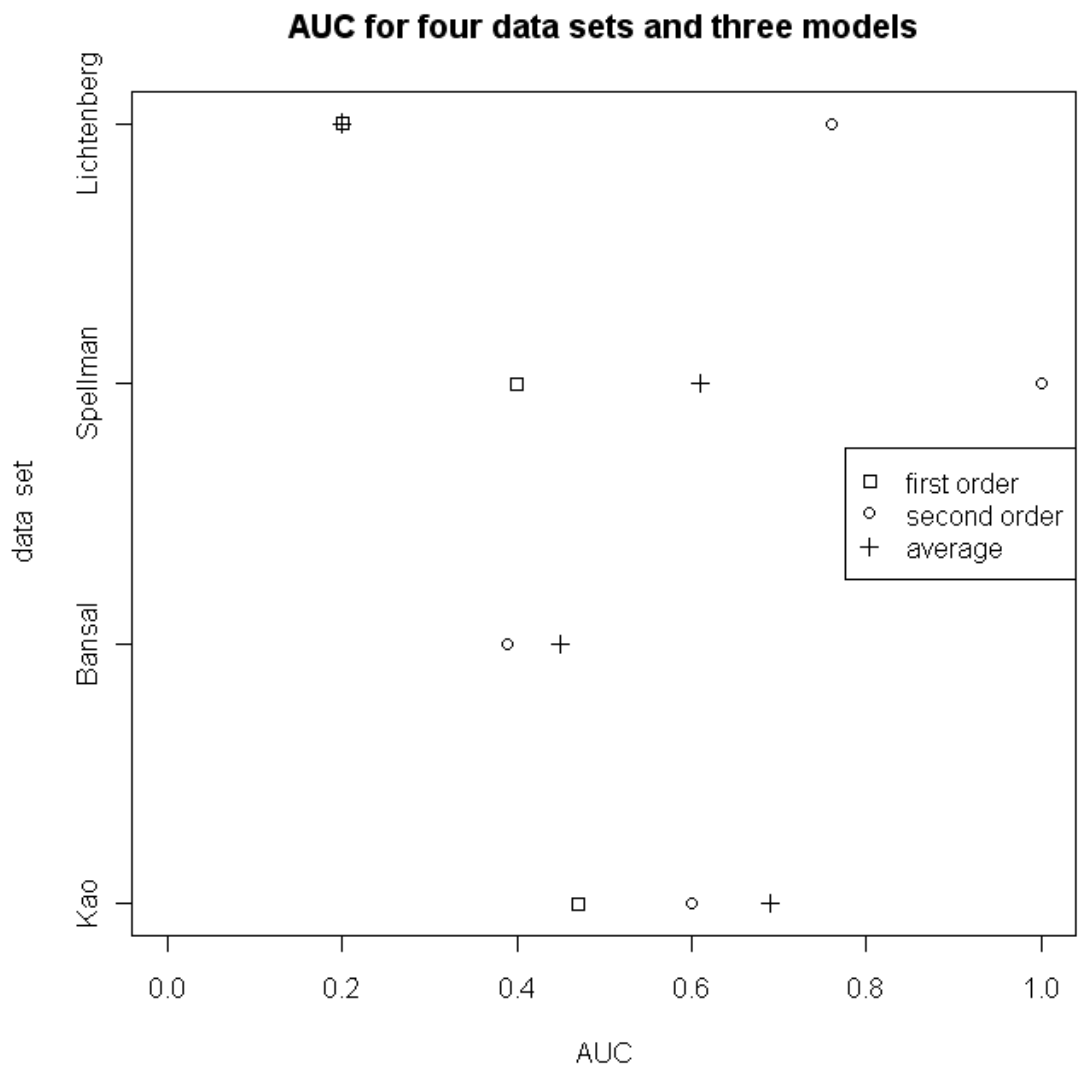

**Figure S16.** The AUC estimates of the four data sets according to first-order, second-order, and average models. The same information appears in Table 1 in the main text.

The numbers of genes or ORFs in the data sets are roughly equal to the genome sizes (5800 to 6000 genes in the yeast genome; 4500 genes in the bacterial genome). Thus, for the purpose of our calculations of the posterior probabilities, the value of $m$ in equations (8) and (12) was set equal to the number of genes or ORFs that are represented in each data set (Tables S3 and S4).

# S3. Sensitivity to the prior distribution

**Implicit priors versus explicit priors**

Although sharing the computational practicality of the BIC approximation, Neyman-Pearson hypothesis testing lacks its coherency, defined as the property that the (implicit) prior distribution stays the same as the sample size increases (Efron and Gous 2001). On the other hand, since the explicit prior distributions of Bayesian methods have been criticized for their subjective or arbitrary selection, the issue of sensitivity of posterior probabilities (14) to the specification of the joint prior distribution of the models and parameters represented by equation (13) will be addressed. The following equations had no bearing on our data analyses, but are included for interested readers.

**Sensitivity to prior probability mass**

Let $J = \arg\max_{j \in \{1,\ldots,m\}} P\left(\alpha_{ij} = 1 \cup \tilde{\alpha}_{ij} = 1 \middle| \mathbf{y}\right)$. Suppose $\omega_i$, the $i$th *relative prior mass*, is the ratio of the prior probability that gene $J$ is the dominant gene regulating of gene $i$ to the prior probability that each other gene $(j \neq J)$ is that regulating gene. Then $P\left(\alpha_{ij} = 1 \cup \tilde{\alpha}_{ij} = 1 \middle| \mathbf{y}\right)$ is approximately

$$\pi_{ij}\left(\omega_i\right) = \left(\hat{\sigma}_{ij}^{-(|\mathcal{T}|-2)n} + \hat{\tilde{\sigma}}_{ij}^{-(|\mathcal{T}|-2)n}\right) C_i \left(I\left(j \neq J\right) + \omega_i I\left(j = J\right)\right) \tag{17}$$

with $\sum_{j'=1}^{m} \pi_{ij'}\left(\omega_i\right) = 1$ redefining $C_i$.

**Sensitivity to prior probability density**

To study the sensitivity to the prior densities of regression coefficients $\beta_{ij}$ conditional on $\alpha_{ij} = 1$, again let $P\left(\alpha_{ij} = 1\right) = 1/m$, i.e., $\omega_i = 1$. Since the BIC approximation

$$P\left(\alpha_{ij} = 1 \middle| \mathbf{y}\right) = \mathrm{BF}_{ij} \chi_i \approx \left(\sup_{\beta_{ij}, \beta_i, \sigma_i} f\left(\mathbf{y} \middle| \alpha_{ij} = 1\right)\right) \chi_i c_i$$

that led to equation (8) has neglected these conditional priors, it is now replaced with a closer approximation that includes them (MacKay 2002):

$$P\left(\alpha_{ij} = 1 \middle| \mathbf{y}\right) \doteq \left(\sup_{\beta_{ij}, \beta_i, \sigma_i} f\left(\mathbf{y} \middle| \alpha_{ij} = 1\right)\right) \phi_{ij},$$

where the maximum likelihood estimate has been substituted for the posterior mode and $\phi_{ij}$ is the Occam factor (Erickson and Smith 1988). The *Occam factor* is, conditional on $\alpha_{ij} = 1$, the ratio of the posterior uncertainty in $\beta_{ij}$, $\beta_i$, and $\sigma_i$ to the prior uncertainty in the same parameters, heuristically speaking. Thus, approximation (8) becomes

$$P\left(\alpha_{ij} = 1 \middle| \mathbf{y}\right) \doteq \phi_{ij} \left(\phi_{ij} + \sum_{j'=1, j' \neq j}^{m} \left(\hat{\sigma}_{ij} \middle/ \hat{\sigma}_{ij'}\right)^{(|T|-1)n}\right)^{-1}.$$

For the purpose of assessing prior sensitivity, define $\Omega_i$, the $i$th *relative Occam factor*, by

$$\phi_{ij} \dot{\propto} I\left(j \neq J\right) + \Omega_i I\left(j = J\right).$$

$\Omega_i$ is the ratio of the Occam factor of gene $J$ to the Occam factor of each gene $j \neq J$. The interpretation is straightforward: if observing the data decreases the regression parameter uncertainty of the $J$th gene more than it decreases the regression parameter uncertainty of each of the other genes, then $\Omega_i < 1$; otherwise, $\Omega_i \geq 1$.

Generalizing to regression coefficients $\beta_{ij}$ and $\tilde{\beta}_{ij}$ conditional on $\alpha_{ij} = 1$ and $\tilde{\alpha}_{ij} = 1$, respectively, let $P\left(\alpha_{ij} = 1\right) = P\left(\tilde{\alpha}_{ij} = 1\right) = \left(2m\right)^{-1}$ and assume that the Occam factor, denoted by $\phi_{ij}$, is the same for models (2) and (3). Then the use of relative Occam factor $\Omega_i$ yields $P\left(\alpha_{ij} = 1 \cup \tilde{\alpha}_{ij} = 1 \middle| \mathbf{y}\right) = \pi_{ij}\left(\Omega_i\right)$; here, $\pi_{ij}\left(\bullet\right)$ is the function defined by equation (17) and $J$ is set identically. Posterior probability $\pi_{ij}\left(\Omega_i\right)$ can be plotted for each data set analyzed; see, e.g., Figure S11.

## S4. Methods of handling missing data

Since the data sets of Section S2 are not complete, we faced what is known in the statistics literature as the missing data problem. We assumed that the probability of missingness was independent of the values of both the missing and observed data; this assumption is termed "Missing Completely At Random" (MCAR). Under the MCAR assumption, we have two options: (i) *available-case analysis*, which uses just the available values of genes in the analysis; or (ii) *imputation*, which first predicts the missing values and then analyzes the complete data (Little and Rubin (2002)). A simulation study demonstrated that the two methods of dealing with missing data show the same performance. We used imputation rather than available-case analysis because we need at least five measurements to apply our method (three time points to find residuals and two measurements to calculate second forward differences) and the Kao *et al.* data has less than five time course measurements for some genes. To impute the missing data, missing values were replaced by the average of the values of their immediate neighbours, and if two or more data are missing side by side, their values were replaced by the average of their neighbouring data points.

## Simulation study

To evaluate the performance of available-case analysis versus our imputation method, we carried out a simulation study. Our simulation not only compared these two methods of dealing with missing data but also checked the accuracy of our algorithm when the noise was increased. In the first step, we generated 10,000 vectors, each with 10 time-course measurements equal to $\sin(t)$ ($t$ is time from 1 to 10) plus normally distributed random deviates. The mean of the normal distribution for each vector was picked randomly from uniform distribution with limits (0.2, 1); the standard deviation of each normal distribution was 0.4. These vectors represent data from a microarray with 10,000 genes; this set of data was kept as complete data. In the second step, we simulated a regulated gene by making it depend on one of the genes from previous step. The regulated gene is simulated according to the first-order model: $x(t+1) - x(t) = \beta x + \beta_0 + \varepsilon(t)$ , where $\beta = 0.7$ , $\beta_0 = 0.5$ , and the noise $\varepsilon(t)$ follows a normal distribution with zero mean and constant standard deviation. The noise standard deviation ($\sigma$) took values in {0, 0.01, …., 0.20} in different simulations. In the third step, we randomly deleted 20% of each data set. In the fourth step, we used equations (8), (12), and (14), to compute the posterior probabilities for complete data. We repeated steps one, two, and four 100 times for each of the 21 values of the standard deviation, thereby simulating the analyses of 2100 microarray data sets. These simulations indicated that the first-order model is able to find the correct regulated gene when $\sigma < 0.06\,\beta$ and is able to find the correct regulating gene more than 50% of time, when $\sigma < 0.14\,\beta$; also, the posterior probability of the first-order model was greater than the posterior probability of the second-order model for all simulations.

To compare the two missing data methods, we repeated steps one through four 50 times for each value of the standard deviation for incomplete data for a total of 1050 simulated microarrays. In step four, posterior probabilities were calculated by using both the imputation and available-case analyses. Our simulations indicated that both methods find the correct gene when the noise is zero and when the first-order model is assumed. When the noise was small ($\sigma \leq 0.04\,\beta$), the available-case method performed slightly better than imputation and vice versa when noise was increased, as seen in Table S5. Results in this table are under the assumption that the first-order model (equation (8)).

| Standard deviation ($\sigma$) | Complete-data (no missing) | Complete-data (imputed) | Incomplete-data (available-case) |
|---|---|---|---|
| 0 | 100% | 100% | 100% |
| 0.01 | 100% | 95% | 100% |
| 0.02 | 100% | 92% | 96% |
| 0.03 | 100% | 90% | 94% |
| 0.04 | 99% | 90% | 82% |
| 0.05 | 95% | 82% | 68% |
| 0.06 | 87% | 80% | 68% |
| 0.07 | 71% | 70% | 58% |
| 0.08 | 64% | 64% | 42% |
| 0.09 | 55% | 40% | 32% |
| 0.10 | 47% | 36% | 22% |
| 0.11 | 32% | 28% | 16% |
| 0.12 | 27% | 20% | 12% |
| 0.13 | 25% | 16% | 12% |
| 0.14 | 23% | 14% | 12% |
| 0.15 | 18% | 12% | 8% |
| 0.16 | 16% | 10% | 8% |
| 0.17 | 11% | 8% | 6% |
| 0.18 | 10% | 8% | 4% |
| 0.19 | 8% | 6% | 4% |
| 0.20 | 6% | 6% | 4% |

**Table S5. Percentage of finding the correct regulating gene for complete and incomplete data.** Our algorithm was applied to simulated complete and incomplete data for different levels of noise. To deal with incomplete data, imputation and available-case methods are used.

## S5. Comparisons to two previous methods

Boolean logic, Bayesian networks, graph theory, additive linear or generalized linear models, differential equations, and stochastic models have all been used to infer regulatory networks. Some of these methods are easy to apply and do not need substantive prior information, whereas the keypad eight others are more demanding but tend to be more accurate. The two methods highlighted here were emphasized by anonymous reviewers; other examples of alternative methods are referenced in the Introduction section of the main text.

## Network identification by multiple regression

Gardner *et al*. (2003) presented a network generated using multivariate regression. They assumed that the cell under investigation is at equilibrium near a steady-state point. They presented a rapid and scalable method to construct prediction model of genes and protein regulatory network without using previous information on the network structure or function. This method is fast and simple, but the algorithm cannot be applied to time course data.

## Inferelator

In recent work, Bonneau *et al*. (2006, 2007) presented a method for deriving genome-wide transcriptional regularity interactions. This method, Inferelator, uses standard regression and model shrinkage techniques to select parsimonious predictive models for gene expression as a function of levels of transcription factors, environmental influences, and interaction between these factors. This method enables simultaneous modelling of equilibrium or steady-state and time-course expression levels. They applied L1-shrinkage to select the predictor and employed tenfold cross validation to determine the optimal value for the shrinkage parameter. They validated their results by further experimentation and also compared them with previous findings. Inferelator has several strengths: (i) since the constructed network is dynamic, it discovers causal relationships by using a time-course expression data between genes; (ii) the modelling of interactions between environmental and transcription factors enables researcher to answer the question of how a simple genetic change or environmental perturbation influences the transcriptional behaviour of a an organism; (iii) the method is not restricted to data with fixed time intervals between measurements. However, to derive accurate results, this algorithm requires extensive quantitative information, which is generally not available, particularly for large regulatory networks. Bonneau *et al*. (2007) used 266 microarray experiments for construction of the network and 147 microarray experiments for validation of predictions. For this large data set, they needed to reduce the dimension of the search space by grouping co-regulated genes into clusters and assuming that the mean expression level of each group of co-regulated genes is influenced by the level of other factors in the system. Investigating all potential regulating genes in the genome was computationally infeasible, so they found putative transcription factors experimentally. This algorithm is useful for investigating the relationship between transcription and environment factors, but a large set of carefully structured data must be available.

Using a simple but flexible model enabled us to consider all genes in the genome as potential regulators and to make inferences on the basis of relatively small data sets. Like Inferelator, the method presented in our study takes advantage of first-order and second-order dynamic models but with stronger assumptions instead of more stringent data requirements. Our algorithm may not need as much data since it relies on a model with only three parameters per gene-gene relationship and is constrained to only one regulator of each gene, whereas Inferelator depends on a model with more parameters and does not share the constraint. Consequently, our method will have greater statistical power to the extent that our assumptions approximate the biological system but can also yield unreliable networks to the extent that our assumptions are violated. As is generally the case, the realism and complexity of statistical models should increase with data availability. Having insufficient data or inappropriately structured data limits options for models in data analysis and may result in misleading assumptions. This underscores the importance of involving experts in statistics and bioinformatics when planning experiments for network reconstruction. For example, the experiment described in Section S1 was designed collaboratively as noted in Section S6.

Our algorithm is presented in the main paper with the restriction of equal sampling times, but it is possible to relax this condition for each gene of interest by fitting a spline or other polynomial to the data and then computing the first and second derivatives of the polynomial instead of taking first and second differences. The posterior probabilities for each gene of interest would then be computed by replacing $\Delta y_{ik}(t)$ in equations (6) and (7) with the first derivative and by replacing $\Delta^2 y_{ik}(t)$ in equation (11) with the second derivative.

## S6. Contributions of each author

DRB made the kinetic model approximations, developed the statistical methods of the main text and Section S3, analyzed the data of Section S1, drafted the manuscript, helped design the experiment of Section S1, and provided guidance for the work of Sections S2, and S4, and S5. ZM helped draft the manuscript, analyzed the yeast and bacteria data sets (Section S2), carried out the simulation study of Section S4, described previous approaches (Section S5), and corrected an error in some plots of Section S1. PH helped draft the manuscript, found the yeast and bacteria data sets, chose the genes of interest for their analyses, and used the results to validate the methods, as described in Section S2. MB helped draft the manuscript and carried out the gene expression profiling, including hybridization, labeling, amplification, quality control, and combination of technical replicates for Section S1. SJL helped draft the manuscript, and designed and executed the sample collection of Section S1, including pooling, centrifugation, and harvesting. NB helped draft the manuscript, designed the experiment, provided guidance in the sample collection, chose the plant genes of interest, and participated in sample harvesting for Section S1.